\documentstyle[sprocl,psfig]{article}

\def\beq{\begin{equation}}
\def\eeq{\end{equation}}
\def\bea{\begin{eqnarray}}
\def\eea{\end{eqnarray}}
\def\ltap{\ \raisebox{-.4ex}{\rlap{$\sim$}} \raisebox{.4ex}{$<$}\ }

\begin{document}

\begin{titlepage}
\noindent
\phantom{a}     \hfill         ZU--TH 10/98            \\
\phantom{a}     \hfill         RU--98--18              \\
\phantom{a}     \hfill         SU--ITP 98--30          \\[3ex] 
\begin{center}

{\bf FERMION MASSES WITHOUT YUKAWA COUPLINGS}      \\[5ex] 
{ F. M.\ BORZUMATI}                                \\[1ex]
{\it Institut f\"ur Theoretische Physik,
     Universit\"at Z\"urich}                       \\
{\it Winterthurerstrasse 190,
     8057 Z\"urich, Switzerland}                   \\[3ex]
{ G. FARRAR, N. POLONSKY}                          \\[1ex]                 
{\it Department of Physics and Astronomy, 
 Rutgers University}                               \\ 
{\it Piscataway, NJ 08854, USA}                    \\[3ex]

{ S. THOMAS}                                       \\[1ex] 
{\it Physics Department, Stanford University, 
     Stanford, CA 94305, USA}                      \\[10ex]
\end{center}
{\begin{center} ABSTRACT \end{center}}
\vspace*{1mm}
{\noindent
Radiatively generated fermion masses 
without tree level Yukawa couplings are re-analyzed within
supersymmetric models.  Special emphasis is given to the  
possible appearance of color and charge breaking vacua.  
Several scenarios in which the radiative mechanism can be 
accomodated for the first, second, and third generation fermion 
masses are presented.  Some of these require a low scale of 
supersymmetry breaking. 
}

\vspace*{2truecm}
{\begin{center} 
{\it Talk presented by F. Borzumati at the Workshop on} \\
{\it Phenomenological Aspects of String Theories}      \\
{\it ICTP, Trieste, October 1997}
\end{center}}

\vfill
\end{titlepage}

\thispagestyle{empty}

\title{FERMION MASSES WITHOUT YUKAWA COUPLINGS}

\author{F. BORZUMATI~\cite{SPEAKER}} 
\address{Institut f\"ur Theoretische Physik, Universit\"at Z\"urich,
 8057 Z\"urich, Switzerland} 

\author{G. FARRAR, N. POLONSKY}
\address{Department of Physics and Astronomy, Rutgers University, 
Piscataway, NJ 08854, USA}

\author{S. THOMAS}
\address{Physics Department, Stanford University, 
Stanford, CA 94305, USA}


\maketitle\abstracts{
Radiatively generated fermion masses 
without tree level Yukawa couplings are re--analyzed within
supersymmetric models.  Special emphasis is given to the  
possible appearance of color and charge breaking vacua.  
Several scenarios in which the radiative mechanism can be 
accomodated for the first, second, and third generation fermion 
masses are presented.  Some of these require a low scale of 
supersymmetry breaking. 
}

\section{Introduction}
\label{sec:intro}

It is often thought that fermion masses cannot be generated
radiatively within supersymmetric models~\cite{RAD}, unless their
generation is due to fermion--sfermion misalignement~\cite{MISAL}. The
radiative mechanism requires large trilinear couplings. These are
generically obtained through the $F$--vacuum expectation value of
spurions which parametrize the breaking of supersymmetry as well as
that of chiral flavour symmetries. The large values required for these
trilinear couplings are believed to produce vacuum instabilities.

Starting from the most general trilinear structure which low--energy
supersymmetric models allow, we reexamine this issue in some
detail. We find that the conventional trilinear couplings, indeed, are
not likely candidates for radiatively generating fermion masses. In
contrast, the ``wrong'' Higgs trilinear couplings, which are absent in
minimal models, may lead to a successful implementation of the
radiative mechanism. For some flavours, a very particular type of
supersymmetry breaking is selected, one in which the scale of breaking,
$M_{SUSY}$, is one or two order of magnitude above the electroweak
scale.  Other options involve mirror fermions at TeV scales.

The paper is organized as follows. In Sect.~2, we discuss the
trilinear couplings which give rise to fermion masses and yukawa
couplings, for which we give explicit expressions in Sect.~3. In
Sect.~4, we give a sufficient condition to avoid unwanted minima. In
Sect.~5, we classify possible scenarios consistent with charge and
color conservation, in which the mechanism of radiative generation of
masses can be implemented. Phenomenological implications for low--energy
and collider physics are discussed elsewhere~\cite{PAPER,MUTALK}.

\section{Classification of operators}
\label{classify}

In the absence of tree--level Yukawa couplings, chiral flavor symmetries 
can be broken by trilinear terms in the scalar potential,
\begin{equation}
V = \sum  m_i^2 \phi_i^2 + 
 \left[ B_{ij} \phi_i  \phi_j + 
        A_{ijk}\phi_i  \phi_j \phi_k + 
        A^\prime_{ijk} \phi_i^\ast \phi_j \phi_k + h.c.
 \right] + \lambda_{ij}\phi_i^2 \phi_j^2.
\label{V}
\end{equation}
The chiral flavor symmetries in the fermion sector are then broken at
the quantum level.  Gauge loops proportional to $A$ or $A^\prime$,
which dress the fermion propagator, generate fermion masses as well as
effective couplings fermion--fermion--Higgs and
fermion--sfermion--Higgsino.

The flavor symmetries of the high--energy theory
can be chosen in such a way to forbid certain fundamental Yukawa
couplings but allow for either operators in the 
superpotential of the type 
$(i) \  Z H \Phi_L \Phi_R /M$,
or operators in the K\"ahler potential 
$(ii) \ ZZ^\dagger H^\dagger \Phi_L \Phi_R /M^3$. 
The chiral superfield $Z = z + \theta^2 F_Z$ 
parametrizes here the supersymmetry breaking sector,
and $\langle F_Z \rangle = M_{SUSY}^2$ signals supersymmetry
breaking at a scale $M_{SUSY}$.
If the scalar component $\langle z \rangle$ vanishes and 
the  auxiliary component $\langle F_Z \rangle$ does not,
then no Yukawa couplings arise but only soft supersymmetry breaking 
trilinear terms $\propto \langle F_Z \rangle^{n} $ in the scalar 
potential. The operators $(i)$ and $(ii)$ lead respectively to 
$A$--  and $A^{\prime}$--type terms, 
\beq 
 (i)  \ A H \phi_L \phi_R,    \hspace*{1truecm} 
 (ii) \ A^\prime H^\ast \phi_L \phi_R , 
\label{aoperators}
\eeq
which are not proportional to any Yukawa couplings.
The  symmetries of the models typically allow
for only one type of operators for a given flavor.
Note that a sufficiently large $A^\prime \sim M_{SUSY}^4/M^3$ 
requires that the supersymmetry breaking scale, $M_{SUSY}$,
and the scale that governs the dynamics in the K\"ahler potential,
$M$, are both relatively low--energy scales. Such a situation could arise,
for example, if there is strong dynamics at  the scale $M$.

\section{Masses and Higgs Couplings}
\label{masscoupling}

The one--loop sfermion--gaugino exchange
which dresses the fermion propagator generates a finite contribution
to the fermion mass. It is given by
\beq
 m_f = -m_{LR}^2
\left\{ \frac{\alpha_s}{2 \pi} C_f  m_{\tilde{g}}  
  I(m^2_{\tilde{f}_1}, m^2_{\tilde{f}_2}, m^2_{\tilde{g}})  
 +      \frac{\alpha^\prime}{2 \pi}  m_{\tilde{B}} 
  I(m^2_{\tilde{f}_1}, m^2_{\tilde{f}_2}, m^2_{\tilde{B}})
\right\},
\label{mass}
\eeq
where $C_f = 4/3, 0$ for quarks and leptons, respectively,
$m_{LR}^2 = A \langle H \rangle$ or 
$A^{\prime} \langle H^\ast \rangle$, and $\tilde{f}_1$, 
$\tilde{f}_2$ are the two mass eigenstates superpartners of 
$f$. The first and second terms correspond to the 
gluino ($\tilde{g}$) and bino ($\tilde{B}$) 
contributions, respectively.  In the second term, corrections
due to possible $\tilde{B}$--$\tilde{W_3}$ mixing are omitted.  The 
function
$I(m^2_{\tilde{f}_1}, m^2_{\tilde{f}_2},m^2_\lambda)$
is such that 
\beq
 I(m^2_{\tilde{f}_1}, m^2_{\tilde{f}_2}, m^2_\lambda) 
\times
\max(m_{\tilde{f}_1}^2,m_{\tilde{f}_2}^2,m^2_\lambda)
\simeq {\cal{O}}(1),
\eeq
where $m_{\lambda}$ denotes generically a gaugino mass. If 
$A$ (or $A^{\prime}$), $m_{\tilde{f}_i}$, and $m_{\lambda}$,
are all of the same order of magnitude, the radiatively 
generated fermion mass is not sensitive to the superpartners
scale and does not vanish even when this is rather 
large. 

In the approximation $\tan \beta \sim 1$ 
(with $\tan \beta$ the ratio of the two vacuum expectation values
$v_2/v_1$), eq.~(\ref{mass}) implies,   
for a typical sfermion mass scale $m_{\tilde{f}}$, 
\bea 
 a \equiv \frac{A}{m_{\tilde{f}}} \ \sim \
 \frac{m_q (M_{\mbox{\tiny weak}})}{(1.5 - 3{\mbox{ GeV}})}
 \hspace*{0.8truecm}& &  {\mbox{ for quarks}}       \nonumber   \\
 \phantom{a \equiv \frac{A}{m}} \ \sim \
 \frac{m_l (M_{\mbox{\tiny weak}})}{(50 - 100{\mbox{ MeV}})}
 \hspace*{0.5truecm}& &  {\mbox{ for leptons}}  
\label{adefvalues} 
\eea 
(and similarly for $a^\prime \equiv A^\prime/m$).
Hence, the maximal magnitude of the trilinear parameters that can be
realized consistently determines which fermion masses can be generated
radiatively.

Effective Yukawa couplings Higgs--fermion--fermion, as well as 
couplings Higgsino--sfermion--fermion,  
are obtained by the corresponding loop diagrams, induced
by the two types of operators $A^\prime H^\ast \phi_L \phi_R$ 
and $AH \phi_L \phi_R$. It is interesting to notice that, in 
general, the Higgsino--sfermion--fermion couplings are 
suppressed with respect to the Higgs--fermion--fermion ones
by factors of order $\alpha_2/\alpha_s $ or 
$\alpha^\prime/\alpha_s $. If fermion masses are generated through 
such a radiative mechanism, large
deviations from the usual hard supersymmetric relations 
among these couplings have to be expected. Details
can be found in Ref.~4. 

It is straightforward to discuss the effective vertex 
Higgs--fermion--fermion with on--shell fields, as in the decay 
$H \to \bar{f}f$, where $H$ is here generically one of the physical 
Higgs states. We denote the relative coupling by $\bar{y}_f$. It 
depends on masses internal and external to the loop which generates it.
In the case of a light Higgs boson $h^0$, when the approximation 
$m_{h^0}/m_{\tilde{f}}, \,\, m_{h^0}/m_{\lambda} \to 0$ 
can be used~\cite{PAPER}, $\bar{y}_f$ has the form 
\beq
 \bar{y}_f = \frac{m_f}{\langle H \rangle}
 \left\{\sin^2 2\theta_{\tilde{f}}
 \left[\frac{1}{2}
  \frac{\sum_i I(m_{\tilde{f}_i}^2, m_{\tilde{f}_i}^2, m_\lambda^2)}
              {I(m^2_{\tilde{f}_1}, m^2_{\tilde{f}_2},m^2_\lambda)}
 - 1\right] + 1\right\}, 
\label{vertex2}
\eeq
where $\sin 2 \theta_{\tilde{f}}$ is the sfermion mixing angle. One
observes that the radiative Yukawa coupling can deviate by a
significant percentage in comparison to the case of a tree--level
fermion mass.  Most importantly, it should be stressed that this
deviation is always an enhancement, which increases with the mass
splitting between the sfermion eigenstates.  This remains true also in
the case of a massive external Higgs boson (see Ref.~4).  Note that
the projecting factors between the physical and interaction Higgs
eigenstates were omitted in eq.~(\ref{vertex2}).  In the case of
$A^{\prime}$--type operators these factors are different than in the
usual case of tree--level couplings.  For $h^0$, this is irrelevant in
the limit of decoupling of the heavy Higgs bosons, which applies to
most of the parameter space. These factors, however, may affect even
further the couplings of heavier Higgs bosons to fermions. They may,
indeed display larger deviations from the couplings relative to
the usual case of tree--level fermion masses.

Other phenomenological consequences of this radiative
scenario can be found in Ref.~4. We concentrate 
in the following on a more fundamental aspect, i.e. 
whether the possibly large trilinear 
scalar operators $AH \phi_L \phi_R$ and 
$A^\prime H^\ast \phi_L \phi_R$ produce vacuum instabilities.

\section{Stability analysis of the scalar potential}
\label{stability}

The low--energy realization of the $F_Z$--spurion 
framework relies
on the presence of substantial dimensionful trilinear couplings $A$ or
$A^{\prime}$ in the scalar potential.  One can typically constrain the
magnitude of such couplings from above by analyzing the vacuum of the
theory.  
It is instructive for our purposes to examine the stability
of the vacuum along an equal field direction corresponding to the
relevant trilinear operator (\ref{aoperators}).

At the tree--level the problem can be partially addressed 
analytically. Following Refs.~6,7, we consider a scalar potential of 
the form
\beq
V(\phi) = m^2 \phi^2 - \gamma\phi^3 + \lambda\phi^4,
\label{potential}
\eeq
where $\phi$ here corresponds to the field along the equal field 
direction $\phi = \phi_{L} = \phi_{R} = H_{\alpha}$, $(\alpha=1,2)$.
More generally, the parameters $m^{2},\gamma,$ and $\lambda$
can depend on various angles, which we ignore here. 
It is then possible to derive a 
condition for color and charge conservation,
\beq
\gamma^2 \leq 4\lambda \, m^2.
\label{color}
\eeq
Condition (\ref{color}) ensures that the deepest minimum along the
equal field direction is at the origin, and hence, the global minimum
of the theory conserves color and charge. It is a sufficient, but not
necessary condition for color and charge conservation. Thus, 
conclusions derived using the sufficient condition (\ref{color})
could, in principle, be weakened. The dimensionful coefficients are
given by
\beq
 m^2 = m_{\tilde{f}_L}^2 + m_{\tilde{f}_R}^2 + m_{H_\alpha}^2,
\label{M2}
\eeq
and 
\beq
\gamma = 2A {\mbox{  (or }} 2A^\prime), 
\label{gamma}
\eeq
where a choice of phase corresponding to the deepest
possible minimum was made. If $m_{\tilde{f}}$ represents 
an average sfermion mass scale, 
$m_{\tilde{f}_L} \sim m_{\tilde{f}_R} \sim m_{\tilde{f}}$,  
then $m^2 \sim 3m_{\tilde{f}}^2$ and $\gamma$ can be expressed 
in terms of the dimensionless variable $a$ ($a^\prime$) 
in (\ref{adefvalues}), 
$\gamma \sim 2 a m_{\tilde{f}}$
(or $\gamma \sim 2a^{\prime}m_{\tilde{f}}$).
It should be noticed that 
the squared Higgs mass contribution in eq.~(\ref{M2}) 
sums over a soft supersymmetry
breaking and a supersymmetry conserving (usually denoted by $\mu^{2}$) 
mass parameters. 
Hence, $m^2 \gg 3m_{\tilde{f}}^2$ is in principle possible 
when a large supersymmetric mass parameter 
$\mu^2 \gg m_{\tilde{f}}^2$ is present. 
Such a large value, however, would correspond to an increased 
degree of fine tuning in the electroweak symmetry breaking 
mechanism. In fact, minimal tuning usually implies 
$|m_{H_2}^2| \sim (1/2)M_Z^2 < m_{\tilde{f}}^2$, and therefore 
$2 m_{\tilde{f}}^2 \ltap m^2 \ltap 3 m_{\tilde{f}}^2$.
Aside from these possible deviations in 
$m_{H_\alpha}^2/m_{\tilde{f}}^2$, 
one can approximate the condition (\ref{color}) with
\beq
|a| \ltap \sqrt{3 \lambda},
\label{colorapprox}
\eeq
and similarly for $a^{\prime}$.  Whether or not this condition is
satisfied depends on the details of the quartic coupling $\lambda$ in
each specific model considered.

The coupling $\lambda$ receives tree--level contributions
from $F$--terms 
\beq 
V_{F} = \sum_i 
\left \vert \frac{\partial W}{\partial \Phi_i} 
\right\vert^2 \sim h^2 \phi^4
\eeq
and from gauge $D$--terms 
\beq
V_{D} = \frac{1}{2} \, g^2 
\left\vert \sum_{i,j} \phi_i^\ast  T^a_{ij} \phi_j \right\vert^2
  \sim \frac{1}{2} \, g^2 ({\mbox{Tr}}Q_i)^2 \phi^4.
\eeq
It would also receive additional 
supersymmetry breaking contributions if a heavy
sector of the theory, which mixes with the light fields, is
integrated out.

It is possible that charge and colour breaking 
global minima appear when values of $|a|$ larger than those 
satisfying condition~(\ref{color}) are required to generate 
radiatively certain masses. In this case, the universe 
may still be on a metastable vacuum with a sufficiently long
lifetime. This somewhat less desirable situation and the 
resulting weaker constraints on $A$, $A^{\prime}$ are 
discussed in Ref.~4.

\section{Possible Models}
\label{sec:models}

Below, we will consider and classify possible contributions
to the quartic coupling $\lambda$. The different contributions
distinguish among the different models.

\subsection{{\em Minimal $A H \phi_L \phi_R $ operator model}}
\label{aminimal}

Such operators correspond to gauge invariant holomorphic operators,
and as such are associated with flat $D$--terms along the equal field
direction, $V_{D} \propto ({\mbox{Tr}}Q_{i})^{2} = 0$.  In the absence
of tree--level Yukawa couplings $V_{F}$ contains only the
supersymmetric mass contribution to the Higgs fields, and hence,
$\lambda = 0$ at tree--level. The trilinear tree--level scalar
potential is unbounded from below. This situation persists at
one--loop where negative quartic couplings 
$\lambda^{OL} \sim -a^{4}/96\pi^{2}$ are generated.

The scalar potential is presumably stabilized at very large field
values due to the physics at those scales.  Nevertheless, the vacuum
cannot be assumed to conserve color and charge, and hence, we are
forced into a metastable vacuum.  As mentioned above, one could 
tolerate such a situation
if the tunneling amplitude to the true vacuum is sufficiently
suppressed.
It would typically require $a \ltap 1$~\cite{PAPER}.

Only very light fermions can be generated radiatively in this type
of models: the electron, the $u$-- and $d$--quark, for which it is 
sufficient to have $a \sim 10^{-3}$.

\subsection{{\em Minimal $A^\prime H^\ast \phi_L \phi_R$ operator
                 model}}
\label{aprimeminimal}

Operators of this type do not correspond to gauge invariant
holomorphic directions and are not necessarily associated with
$D$--flatness. In particular, non--flat is the 
hypercharge $D$--term since $Y(H) = (Y(\phi_L) +Y(\phi_R))$. We 
obtain in this case (independently of flavor labels):
\begin{equation}
\lambda  \sim \frac{1}{2} \, g^{\prime 2} \sim 0.06.
\label{lambda1}
\end{equation}
The potential is now bounded from below. Substituting (\ref{lambda1})
in condition (\ref{colorapprox}) gives $a^{\prime} \ltap  0.4$.

Thus, the $c$-- and $s$--quark masses, which require respectively 
$a^\prime = 0.2$--$0.5$ and $0.1$, can be easily accommodated in this
model. 

\begin{table}[t]
\caption{The quartic coupling along the equal field direction
in the different scenarios.}
\label{table:t1}
\vspace{0.2cm}
\begin{center}
\footnotesize
\begin{tabular}{|l|c|l|l|}
\hline
 Model & $\lambda$ & Limits & Comments \\
\hline
 Minimal $A$ & $\sim -(1/96\pi^{2})a^{4}$ & $a\, \  \ltap 1$ 
                                          & a metastable vacuum        \\
 Minimal $A^{\prime}$ & $g^{\prime 2}/2$  & $a^{\prime} \ \ltap 0.4$
                                          &                            \\
 Hybrid $h_{t} -A^{\prime}$ & $h_{t}^{2}$ & $a^{\prime} \ \ltap \sqrt{3}$ 
                                          & relevant for $b$           \\
 Mirror matter& $h^2 [{\widetilde{m}}^2/({\widetilde{m}}^2+\mu^2)]$ 
                                          & $a$,$a^{\prime} \ltap$ a few 
                                          & assumes a multi--TeV scale \\
\hline
\end{tabular}
\end{center}
\end{table}

\subsection{{\em A hybrid model $W\sim h_t H_2 QU$ and 
                 $V\sim A^\prime H_2^\ast QD$}}
\label{hybrid}

Here we will consider a specific example motivated by the sharp
distinction between the $t$--quark Yukawa couplings $h_{t} \sim 1$ and
all other low--energy Yukawa couplings $h_{f} \ll 1$ (for 
$\tan\beta \ll 50$). The presence of the tree level supersymmetric 
operator $h_{t}H_{2}QU$ carries important consequences for the scalar
potential along the equal field direction associated with the 
supersymmetry breaking operator $A^\prime H_2^\ast QD$, which could be
a source for the $b$--quark mass.  (Note that 
we do not distinguish in our notation
between a standard matter chiral superfield and its scalar 
component.)

The $F$--terms contains the  quartic term 
$|\partial W/\partial U|^2 = h_t^2 H_2^2 Q^2$ and hence,
\beq
\lambda  \sim  h_t^2 +\frac{1}{2} \, g^{\prime 2} \sim h_t^2 \sim 1.
\label{lambda2}
\eeq
In the case of the $b$--quark mass one needs 
$a^{\prime} \ltap \sqrt{3}$, which can be easily accommodated in 
this model.  Of ${\cal O}(1)$ is also the $a (a^\prime)$ needed for
the muon mass. Such a large value excludes the radiative generation of
this mass in the models described in Secs.~5.1 and~5.2. A mechanism 
similar to 
that described in this section can work, if there exist a term in the
superpotential involving second generation lepton fields with a
large coupling. This can then play the same stabilizing role that 
$h_t$ has in the case of the $b$--quark mass, as it will be 
discussed in the next section.

\subsection{{\em Models with mirror matter}}
\label{mirror}

Vector--like (mirror) pairs of chiral superfields exist in many
extensions of the standard model near or above the weak scale.
Often such fields are expected to have the same 
transformation properties under the SM gauge group as the SM fields.
They may transform differently under additional symmetries, in
particular, flavor symmetries. 

It is natural to expect some mixing between the two sectors, which
lead to a lower bound on the mass scale of the exotic matter. The new
scale could be set by either supersymmetric mass terms for the
vector--like pairs or by soft supersymmetry breaking parameters.  It is
also reasonable to expect that the mirror matter fields transform
differently than the SM fields under the flavor symmetries, and
therefore Yukawa operators that mix SM and exotic fields may be
allowed with large couplings. Clearly, the role played by the SM
superfield $U$ in the previous example can be played in this case by
an exotic field. Furthermore, there are no restrictions, in this case,
on the flavor labels or on the type of operator.

Motivated by recent discussions of decoupling in supersymmetric 
models~\cite{CKN}, we will assume, for simplicity, that all soft
supersymmetry breaking ${\widetilde{m}}^{2}$ and supersymmetry conserving 
mass parameters $\mu^{2}$ for the exotic fields (and perhaps for some SM
fields) are multi TeV parameters.  After integrating out the heavy
fields one obtains for the light fields
\beq
\lambda  = h^2
 \left( \frac{\widetilde{m}^2}{\widetilde{m}^2 + \mu^2} \right),
\label{lambda3}
\eeq
where $h$ is the relevant Yukawa coupling between the Higgs, SM and
exotic heavy superfields, $h H \phi_{SM} \phi_{heavy}$.  The 
integration of the heavy fields leads to a
supersymmetry breaking contribution to $\lambda$.  (In fact, one has
$\lambda = 0$ in the supersymmetric limit.)  The usual $h^{2}$ term is
now modified by an {\em a priori} arbitrary factor of ${\cal{O}}(1)$.
Since a few of these ${\cal{O}}(1)$ contributions may be present, it
is possible to have in this case $a, \, a^{\prime} \ltap$ a few.  

The muon mass can then be generated radiatively in models of this
type. The far--reaching phenomenological consequences 
for the process $\mu^+ \mu^- \to H \to f \bar{f}$ and for the 
muon magnetic moment are discussed in Refs.~4,5.

\section{Conclusions}
\label{concl}

We have re--analyzed the problem of radiative generation of fermion
masses through trilinear soft operators. These are obtained from
holomorphic and/or non--holomorphic operators in which a spurion field
acquires a vacuum expectation value only in the $F$--component,
breaking simultaneously supersymmetry and chiral fermion
symmetries. In general, large values of these trilinear couplings are
needed for the radiative generation of second and third generation
fermion masses.
We have found several consistent scenarios in which fermion masses for
light and heavy flavours could be generated consistenly and with
stable vacua. Some of these scenarios seem to point to a
scale of supersymmetry breaking not far above the electroweak scale.

\section*{Acknowledgments}
This work was supported by Schweizerischer Nationalfonds, by 
the US National Science Foundation grant PHY--94--23002,
and by Stanford University through the Fredrick E. Terman 
Fellowship.

\end{document}